\documentclass{optica-article}

\journal{opticajournal} 

\articletype{Research Article}

\usepackage{lineno}
\usepackage{siunitx}
\usepackage[version=4]{mhchem}

\begin{document}

\title{Broadband and flat-baseline dual-comb cavity mode dispersion spectroscopy using fiber-based frequency combs}

\author{Sho Okubo,\authormark{1,*} Ken Kashiwagi,\authormark{1} Koji Hashiguchi,\authormark{1} and Hajime Inaba\authormark{1}}

\address{\authormark{1}National Metrology Institute of Japan (NMIJ), National Institute of Advanced Industrial Science and Technology (AIST), 1-1-1 Umezono, Tsukuba 305-8563, Japan}

\email{\authormark{*}sho-ookubo@aist.go.jp} 


\begin{abstract*} 
We demonstrate broadband dual-comb cavity mode dispersion spectroscopy using mode-locked erbium-fiber frequency combs and a length-tunable optical cavity. The flat baseline and low-noise characteristics of cavity-mode-dispersion spectroscopy, combined with the broadband coverage of fiber frequency combs, provide highly precise spectral profiles that are difficult to obtain with conventional dual-comb spectroscopy. We measured the entire $\nu_1 + \nu_3$ band of acetylene in the \qty{1.5}{\um} region. The simultaneous fitting of multiple transitions agreed well with the measured spectra, yielding a relative standard deviation of \qty{0.27}{\%} for the retrieved acetylene pressure and a spectral fluctuation corresponding to an absorption coefficient of \qty{1.4e-6}{cm^{-1}}. These results reveal the high precision achievable with the present method. Further sensitivity improvements are expected with a higher-finesse cavity.
\end{abstract*}


\section{Introduction}
The optical frequency comb \cite{Hall2006, Hansch2006} was developed as a tool for optical frequency measurement in practical realization of the difinition of the meter \cite{FelderR2005} and the development of next-generation atomic clocks, and has contributed to the advancement of various scientific fields such as precision spectroscopy \cite{PicqueN2019}, distance and shape measurement \cite{Minoshima2000, OhJ2005}, low-noise microwave generation \cite{RamondT2002, EndoM2018}, and astronomy \cite{MurphyM2007, McCrackenR2017}. In precision spectroscopy, several methods that employ the broad bandwidth, accuracy, and controllability of the comb have been proposed and studied for environmental, medical, and industrial applications. Performance metrics in spectroscopy include sensitivity, spectral resolution and accuracy, wavelength range, and measurement speed. For example, by offset-locking a CW laser to an atomic-clock-referenced comb and sweeping its repetition rate, we can sweep the CW frequency laser near absorption lines and obtain an absorption spectrum with the highest-level spectral resolution, signal-to-noise ratio (SNR), and frequency accuracy. This has been used in several studies regarding the shapes of molecular absorption lines \cite{CichM2013}, the verification of pressure effects appearing in absorption spectra \cite{OkuboS2023}, and the high-precision measurement of thermodynamic temperature from Doppler widths \cite{Yamada2009, GattiD2013}. In contrast, there are numerous studies that use the frequency comb directly as a spectroscopic light source, separating and observing each comb mode arranged at equal intervals across a wide range in the frequency domain. These methods allow the simultaneous observation of the way in which each comb mode is absorbed by the sample, thus enabling a high-resolution, broadband, and high-precision spectrum to be measured in a short time. Methods for separating and detecting each comb mode include two-dimensional dispersive spectroscopy using high-resolution dispersive elements such as a virtually imaged phase array (VIPA) \cite{Diddams2007} or an immersion grating \cite{IwakuniK2019}, and Fourier transform spectroscopy (FTS) using a Michelson interferometer \cite{Mandon2009} or dual-comb spectroscopy (DCS) \cite{Schiller2002, KeilmannF2004, Coddington2008}. These direct comb spectroscopy techniques have been used for numerous spectroscopic applications, such as studies of intermediates and reaction dynamics in chemical reactions via trace radical molecular spectroscopy (VIPA-type two-dimensional dispersive spectroscopy) \cite{Fleisher2014}, spectroscopic diagnostics in combustion fields (Michelson interferometer) \cite{AlrahmanC2014}, and livestock greenhouse gas emission monitoring (DCS) \cite{WeerasekaraC2024}.

Among these, DCS is the most actively researched. DCS is a type of FTS that uses two optical frequency combs. Unlike conventional FTS, which employs incoherent light sources and a Michelson interferometer, DCS acquires interferograms by utilizing the difference in pulse repetition rates between two combs, without any mechanically moving parts in the interferometer. As a result, it can generate large delays in a short time without mechanical instability, enabling the acquisition of a high-precision, high-resolution, broadband spectrum in a brief period. Compared with two-dimensional dispersive spectroscopy, DCS offers several advantages: it does not require expensive imaging devices, is robust against optical system fluctuations with minimal variation in the spectral axis, and achieves high maximum resolution, allowing mode-resolved spectra even when the comb’s repetition frequency is of the order of tens of MHz across the entire comb bandwidth. In DCS, synchronizing the comb frequency to a reference yields high accuracy as regards the horizontal axis (frequency) of the observed spectrum. Using a commercial Rb atomic clock (with a relative accuracy of about \num{e-11}) as a reference, it is possible to determine transition frequencies from Doppler-limited molecular spectra. Furthermore, the resolution is limited by the repetition frequency, which is typically higher than that of conventional FTS. By interpolating the spectrum, the resolution can be further enhanced so that it is comparable to the comb mode linewidth \cite{Okubo2015d}. These characteristics indicate that the horizontal axis performance of the spectrum is excellent. This high accuracy in the horizontal axis minimizes the fluctuations in the vertical axis (amplitude, power) caused by frequency instability. However, in practical DCS, unwanted interference (etalon effects) in the spectrometer and the frequency dependence of the comb mode power limit the vertical axis accuracy of the obtained spectrum, resulting in spectral distortion. Such distortion can cause discrepancies between the measurement and the model used to determine the sample concentration or temperature, resulting in uncertainty that limits measurement accuracy. To eliminate spectral distortion, normalization using a background spectrum measured without the sample is often employed. However, due to etalon effects and temporal variations in the comb spectrum \cite{IrimatsugawaT2021}, such normalization cannot completely remove spectral distortion. In cases where a background spectrum cannot be directly obtained, such as in open-air measurements \cite{Rieker2014, WaxmanE2017, HanJ2024}, alternative methods for estimating the background spectrum have been proposed and demonstrated. One example is to divide the measured spectrum into frequency windows (about 100 GHz each), fit each segment with a sample absorption lineshape function and a polynomial baseline, and then stitch together the resulting polynomial baselines \cite{WaxmanE2017}. Another example is to separate the components reflecting the background and the free induction decay, which reflects sample absorption, in the interferogram to estimate the background spectrum \cite{ColeR2019}. These methods also have the advantage of being less affected by temporal spectral variations. On the other hand, when the spectral power dependence on the optical frequency, which fluctuates over time, is comparable to or smaller than that due to sample absorption, it becomes difficult to distinguish between absorption and background components. This can potentially reduce the accuracy of background separation. As an alternative approach to spectral normalization and background estimation, photoacoustic spectroscopy, which provides background-free molecular absorption spectra, has also been incorporated into dual-comb spectroscopy and combined with an optical cavity for sensitivity enhancement, yielding high-quality spectra \cite{WangZ2024b}. Thus, separating absorption and background components remains a major challenge in DCS, as well as with other spectroscopic methods.

Cavity mode dispersion spectroscopy (CMDS) \cite{CyganA2013} has been proposed as an alternative approach to suppressing the spectral distortion caused by background components such as the etalon effect of the spectrometer and the frequency dependence of the comb mode power. This approach utilizes the cavity resonance frequency shift induced by a refractive index change in the sample inside the cavity. CMDS is a spectroscopy technique based on frequency measurement and is less susceptible to the etalon effect of the spectrometer or the frequency dependence of the comb mode power, thus offering the advantage of reducing spectral distortion in the observed spectrum. CMDS using a CW laser as the light source and the determination of the transition frequency have been demonstrated \cite{CyganA2016}. CMDS using a dual-comb system based on modulator-type frequency combs has also been reported \cite{CharczunD2022}. The combination of DCS and CMDS is promising for eliminating background components in absorption spectra, specifically the effects of the etalon and the frequency dependence of comb mode power, and is expected to enable the observation of spectra with minimal distortion over a wide wavelength range. However, in previous studies, the wavelength range was limited due to the use of modulator-type combs, and the acquisition of low-distortion spectra spanning an entire molecular rovibrational band has not yet been achieved. Obtaining such a broadband spectrum with suppressed background distortion is beneficial for a wide range of analyses. For example, it is expected to improve the accuracy of the spectral-model-based determination of gas concentrations \cite{WaxmanE2017}, isotope ratios \cite{ParriauxA2022}, and temperature \cite{KloseA2016, ShimizuY2018}. In addition, energy levels, transition strength, and nontrivial shifts and the broadening of the transition spectrum induced by collisional effects can be validated with high precision by comparison with an ab initio molecular science calculation \cite{HartmannJ2013}.

In this study, we combined CMDS with DCS using an erbium-doped-fiber-based mode-locked laser as the comb source. Mode-locked-laser-based combs can easily provide a broader spectral bandwidth than modulator-based combs. As a result, we successfully observed the dispersion spectrum covering the entire $\nu_1 + \nu_3$ band of acetylene (\ce{^{12}C2H2}) in a single measurement, achieving a broader spectral range than in a previous study \cite{CharczunD2022}. By comparing the obtained spectra with those obtained with conventional DCS, we demonstrated that the background components caused by fringes were significantly suppressed across the entire rovibrational band without spectral normalization. Furthermore, we constructed a model equation for the cavity-mode-dispersion (CMD) spectrum with spectroscopic parameters from the HITRAN database \cite{GordonI2022} and fitted it to the observed spectra using a least-squares method to determine the \ce{C2H2} pressure. The resulting pressure values agreed with those obtained with a vacuum gauge within the measurement uncertainty, confirming the validity of both the spectral measurement and analysis methods. Additionally, we estimated the minimum detectable absorption coefficient (MDAC) \cite{OKeefeA1988} based on the deviation of the CMD spectra to evaluate the detection sensitivity of our method. These comprehensive evaluations enabled a direct comparison of our method with conventional techniques, demonstrating the effectiveness of the proposed approach.

\section{Concept and experimental setup}

\begin{figure}[htbp]
\centering\includegraphics[width=9cm]{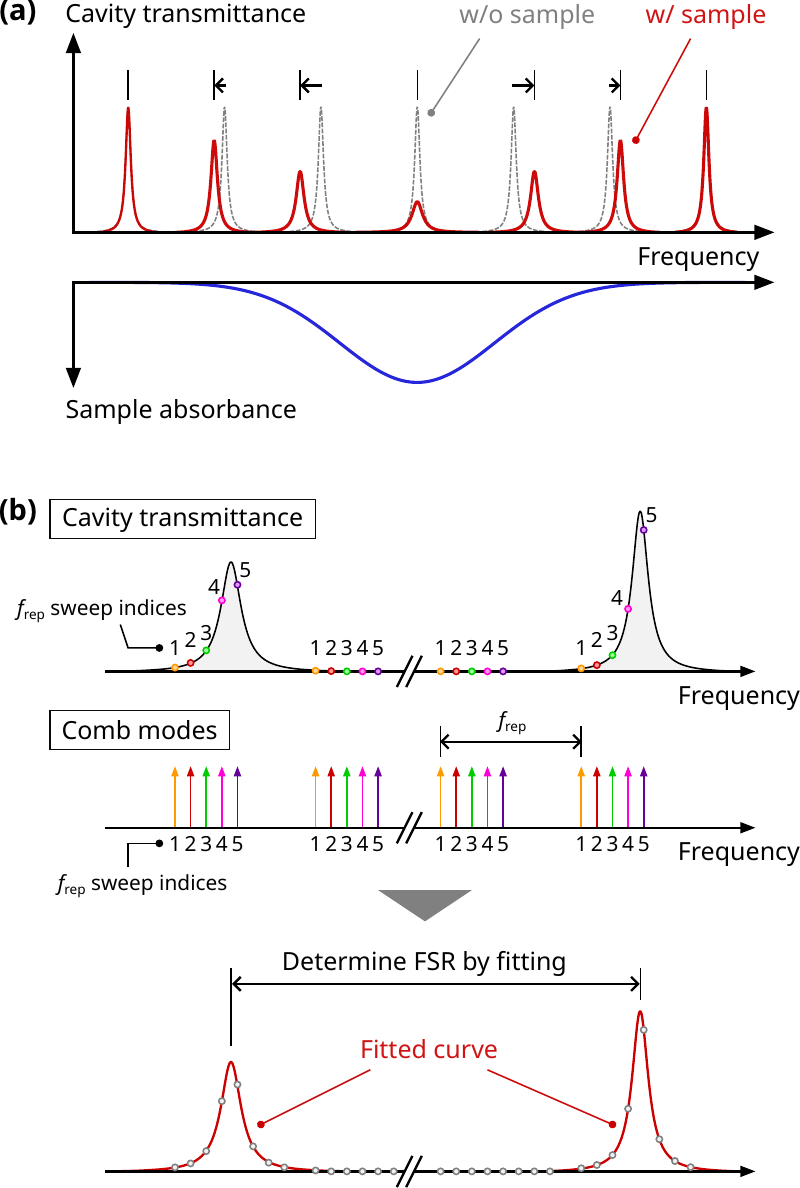}
\caption{(a) Concept of CMDS. (b) Procedure for CMDS with dual-comb spectrometer.}
\label{fig-1}
\end{figure}

Figure \ref{fig-1}(a) illustrates the concept of CMDS. With a Fabry-Pérot cavity, when the round-trip distance equals an integer multiple of the wavelength of light, the following resonance condition
\begin{equation}
\nu\left(m\right)=m\frac{c}{2nL} \label{eq-1}
\end{equation}
is satisfied, and the incident light with frequency $\nu(m)$ resonates within the cavity. Here, $c$ is the speed of light, $L$ is the cavity length (distance between the cavity mirrors), $n$ is the refractive index of the medium between the mirrors, and $m$ is the cavity mode number. When $m$ differs by one, the resonant frequency of the light differs by the free spectral range (FSR) given by $c/(2nL)$. Therefore, the transmission spectrum of the optical cavity consists of a large number of transmission modes spaced by the FSR. The refractive index $n$ depends on the medium between the mirrors, namely on the electrical susceptibility of the sample introduced into the cavity. Near the frequency where the sample inside the cavity absorbs light, the refractive index in Eq. (\ref{eq-1}) changes, resulting in a shift in the resonant frequency. In this case, the cavity transmittance $T(\nu)$ is expressed as follows \cite{CyganA2015}:
\begin{equation}
T\left(\nu\right) = \frac{{T_\mathrm{m}}^2e^{-\alpha\left(\nu\right)L}}{\left(1-R_\mathrm{m}e^{-\alpha\left(\nu\right)L}\right)^2+4R_\mathrm{m}e^{-\alpha\left(\nu\right)L}\sin^2{\left(\frac{\pi\nu}{\xi\left(\nu\right)}\right)}} \label{eq-2}
\end{equation}
Here, $R_\mathrm{m}$ and $T_\mathrm{m}$ are the reflectance and transmittance of the cavity mirrors, respectively, and $\alpha(\nu)$ is the absorption coefficient (absorbance per unit length) due to resonance with the sample in the cavity. The parameter $\xi(\nu)$ corresponds to the FSR considering the interaction between the sample and light in the cavity. Note that $\xi(\nu)$ is typically smaller than the linewidth of the sample absorption lines. The functions $\alpha(\nu)$ and $\xi(\nu)$, respectively, are given by
\begin{align}
\alpha\left(\nu\right) &= \frac{cSp}{kT}V_\mathrm{abs}\left(\nu-\nu_0\right) \label{eq-3}\\
\xi\left(\nu\right) &= \nu_\mathrm{FSR,vac}\left[n_0-\frac{c^2Sp}{4\pi\nu_0kT}V_\mathrm{disp}\left(\nu-\nu_0\right)\right]^{-1} \label{eq-4}
\end{align}
where, $\nu_0$ is the transition frequency of the sample, $S$ is the spectral line intensity of the transition, $p$ is the sample pressure, $k$ is the Boltzmann constant, $T$ is the sample temperature, $V_\mathrm{abs}(\nu)$ and $V_\mathrm{disp}(\nu)$ are the area normalized absorption- and dispersion-type Voigt functions, and $\nu_\mathrm{FSR,vac} = c/(2L)$ is the FSR of the vacuum cavity. $n_0$ is the refractive index of the sample medium in the absence of any interaction with the light, and $n_0 \simeq 1$ in this study. When the cavity length $L$ is constant, Eq. (\ref{eq-2}) shows that the interaction between the sample and light causes a reduction in the transmittance and a shift of the cavity resonant frequency. Therefore, the absorption strength of the sample can be determined from the variation of in the cavity resonant frequency. In practice, however, the cavity resonant frequency also varies due to the fluctuation of the cavity length $L$ caused by thermal expansion and mechanical vibration. To suppress these effects, the cavity resonant frequency is stabilized to a reference CW laser in a frequency region without sample absorption. By stabilizing L, it becomes possible to detect only the change in the cavity resonant frequency induced by sample absorption.

Figure \ref{fig-1}(b) shows the dual-comb CMDS (DC-CMDS) procedure described in this study. With this method, the linewidth of each cavity transmission mode is narrower than the comb repetition frequency frep. To observe this, a spectral interleaving technique \cite{Okubo2015d} is used, in which the comb-mode frequencies are swept in small, discrete steps to obtain the cavity-mode transmission spectrum. Figure \ref{fig-1}(b) shows how comb-mode frequencies are discretely swept by changing frep. The indices 1-5 in Fig. \ref{fig-1}(b) represent the first five setpoints of a total of ten discrete $f_\mathrm{rep}$ setpoints for spectral interleaving. For clarity, the corresponding comb-mode frequency displacement is shown as $f_\mathrm{rep}/10$ in Fig. \ref{fig-1}(b). Comb-mode frequencies are shifted accordingly between adjacent $f_\mathrm{rep}$ setpoints. The comb modes and the data points of the cavity transmittance spectrum assigned the same index correspond to the comb-mode group present at a given frep setting and the spectrum recorded by that group. The cavity transmittance spectrum is obtained by recording the spectrum at each $f_\mathrm{rep}$ setpoint and stitching the spectra together. The obtained spectrum data are then fitted to a theoretical model for the cavity transmittance spectrum using a least-squares method to determine the center frequency of each cavity mode. Finally, the difference between the center frequencies of adjacent cavity modes is calculated, yielding the one-dimensional CMD spectrum \cite{CyganA2015}.

\begin{figure}[htbp]
\centering\includegraphics[width=13cm]{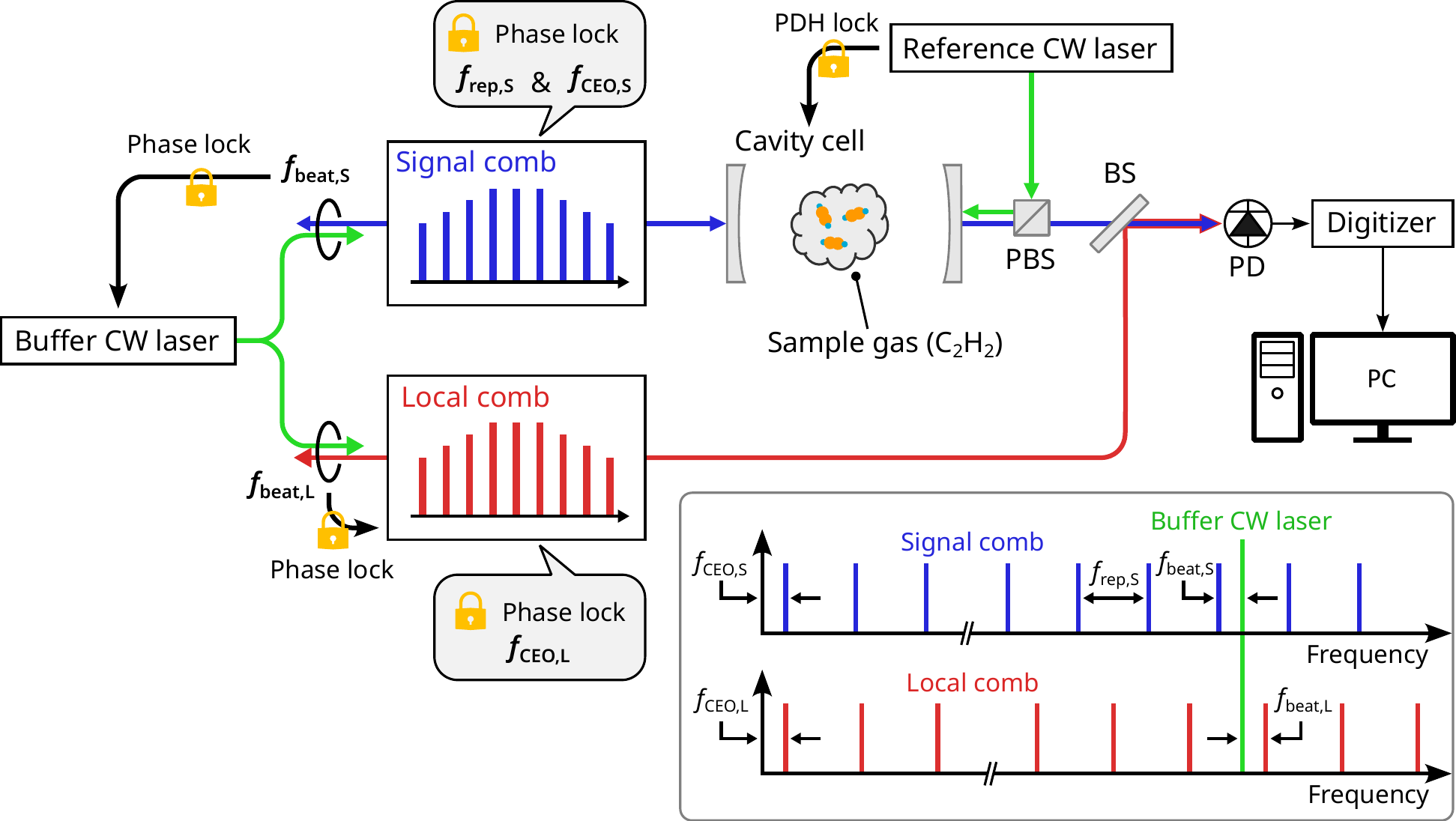}
\caption{Experimental setup. PBS: polarizing beam splitter, BS: beam splitter, PD: photodetector, PDH lock: Pound-Drever-Hall lock.}
\label{fig-2}
\end{figure}

Figure \ref{fig-2} shows the experimental setup used in this study. The dual-comb system is based on the one used in our previous studies \cite{Okubo2015d, Okubo2015a, IwakuniK2016b, OkuboS2017, ShimizuY2018}. A brief explanation follows: Two optical frequency combs (referred to as the signal and local combs) are generated from mode-locked Er-doped fiber lasers. The signal and local combs are set at slightly different repetition frequencies, with $f_\mathrm{rep,S} \simeq \qty{48.35}{MHz}$ and $f_\mathrm{rep,L} = f_\mathrm{rep,S} - \Delta f_\mathrm{rep}$, respectively. The beam of the signal comb passes through a cavity integrated with the sample cell (cavity cell) and is overlapped with the beam of the local comb at a beam splitter. The interference signal between the signal and local combs (interferogram) is then detected with a photodetector. The local comb serves as the local oscillator of a multi-heterodyne technique for reading the spectroscopic signal transferred to the signal comb. The detected interferogram is sampled by a digitizer and sent to a PC, where the spectrum is computed using a fast Fourier transform algorithm. In addition, a reference CW laser is counter-propagated through the cavity cell relative to the signal comb via a polarizing beam splitter. The reference CW laser is stabilized to a highly stable optical cavity with an ultra-low thermal expansion glass spacer and used as a reference for the length of the cavity cell.

The cavity cell is an optical cavity integrated with a sample gas cell constructed as a vacuum chamber. To avoid limiting the finesse due to losses at the chamber windows, the cavity cell is designed such that the Fabry-P\'{e}rot cavity is placed inside the vacuum chamber \cite{HashiguchiK2024}. Bellows are attached to both ends of the chamber, and mirror holders and chamber windows are mounted at the ends of the bellows. Each mirror holder allows tilt adjustment via a kinematic mount and translation along the optical axis via a translation stage. By combining the bellows and mirror holders, it is possible to adjust the optical axis of the cavity mirrors and perform coarse tuning of the cavity length while keeping the sample gas sealed inside the chamber. Furthermore, to control the cavity length precisely, one of the mirrors is mounted on a piezoelectric transducer (PZT) in the mirror holder. These cavity length control mechanisms are based on a previously reported design \cite{NakamuraK2023}, and further details will be reported in another paper. The cavity length is approximately \qty{51}{cm}, corresponding to an FSR of $\sim \qty{290}{MHz}$. It is designed such that the FSR can be adjusted to exactly six times the $f_\mathrm{rep,S}$, so that every sixth signal comb mode is transmitted through the cavity. This allows the sweep range of the comb-mode frequencies to be limited to a narrow region around the cavity resonances. The cavity mirrors used in this study were designed for the \qty{1480}{nm} to \qty{1650}{nm} wavelength range, with a reflectance of $\sim 0.99$ (corresponding to a finesse of $\sim 300$, a cavity mode full width at half maximum (FWHM) of $\sim\qty{0.9}{MHz}$) and a group delay dispersion of less than \qty{0.15}{fs^2}. Although mirrors with moderate reflectance were used for the demonstration experiment, mirrors with higher reflectance can be used for measurements requiring greater sensitivity. The cavity cell is equipped with flexible tubing and valves for use in installing and sealing the sample gas, as well as a vacuum gauge for pressure measurement. In this study, \ce{C2H2} at \qty{10}{Pa} was used as the sample.

Below, we describe the procedure for the frequency stabilization of the cavity and combs, as well as the spectrum measurement of the comb transmitted through the cavity. First, the repetition and carrier-envelope offset frequencies of the signal comb are phase-locked to $f_\mathrm{rep,S}$ and $f_\mathrm{CEO,S}$, respectively. Next, the cavity length is scanned to observe the transmission peaks of the reference CW laser and signal comb. Then the cavity length is adjusted so that the separation between the two peaks is minimized. Under this condition, $f_\mathrm{rep,S}$ is adjusted to make the two peaks coincide. After that, the cavity mode is stabilized to the frequency of the reference CW laser using the Pound-Drever-Hall technique \cite{Drever1983}.

The local comb is phase-locked relative to the signal comb via an intermediate CW laser. The frequency relationships between the signal comb, the local comb, and the intermediate CW laser are shown in the Fig. \ref{fig-2} inset. The intermediate CW laser is phase-locked such that the beat frequency between the intermediate CW laser and one mode of the signal comb is set at $f_\mathrm{beat,S}$. Next, the beat frequency between the intermediate CW laser and one mode of the local comb is phase-locked to $f_\mathrm{beat,L}$, and the carrier-envelope offset frequency of the local comb is phase-locked to $f_\mathrm{CEO,L}$. By appropriately selecting the comb mode numbers used to detect the beat signals with the intermediate CW laser, as well as adjusting $f_\mathrm{CEO,S}$, $f_\mathrm{beat,S}$, $f_\mathrm{CEO,L}$, and $f_\mathrm{beat,L}$, the conditions for coherent averaging \cite{Coddington2009b} at a desired repetition frequency difference $\Delta f_\mathrm{rep}$ can be satisfied. This enables the application of DCS with a resolution that is not limited by the time window length when recording the interferogram \cite{Okubo2015d, Yasui2015a}. Under this condition, DCS can resolve spectral features that are narrower than the the comb repetition frequency, thus allowing the observation of cavity transmission modes with a linewidth of $\sim \qty{0.9}{MHz}$ \cite{Okubo2015d}. In this study, $\Delta f_\mathrm{rep}$ was set at \qty{81}{Hz}, corresponding to a measurable frequency range of approximately \qty{14}{THz} without aliasing.

Using the phase-locked combs described above, we measure the cavity transmission spectrum by DCS, and detect changes in the cavity resonant frequencies. The repetition frequency of the signal comb is selected at several discrete values around $f_\mathrm{rep,S}$, where the cavity-transmitted power of the signal comb is maximized, and a spectrum measurement with DCS is performed at each repetition frequency. In this study, $f_\mathrm{rep,S}$ was swept over 39 setpoints with a step size of \qty{0.05}{Hz}, over a total range of \qty{1.9}{Hz}. This corresponds to an optical (comb-mode) frequency step of $\sim \qty{200}{kHz}$ and a total range of $\sim \qty{7.7}{MHz}$ (\qty{16}{\%} of $f_\mathrm{rep,S}$). The interferogram was recorded with an average of 3000 acquisitions at each $f_\mathrm{rep,S}$ setpoint. The spectra obtained by Fourier transforming these interferograms were then stitched together to obtain the cavity transmission spectrum.

\section{Results and analysis}

\begin{figure}[htbp]
\centering\includegraphics[width=8.5cm]{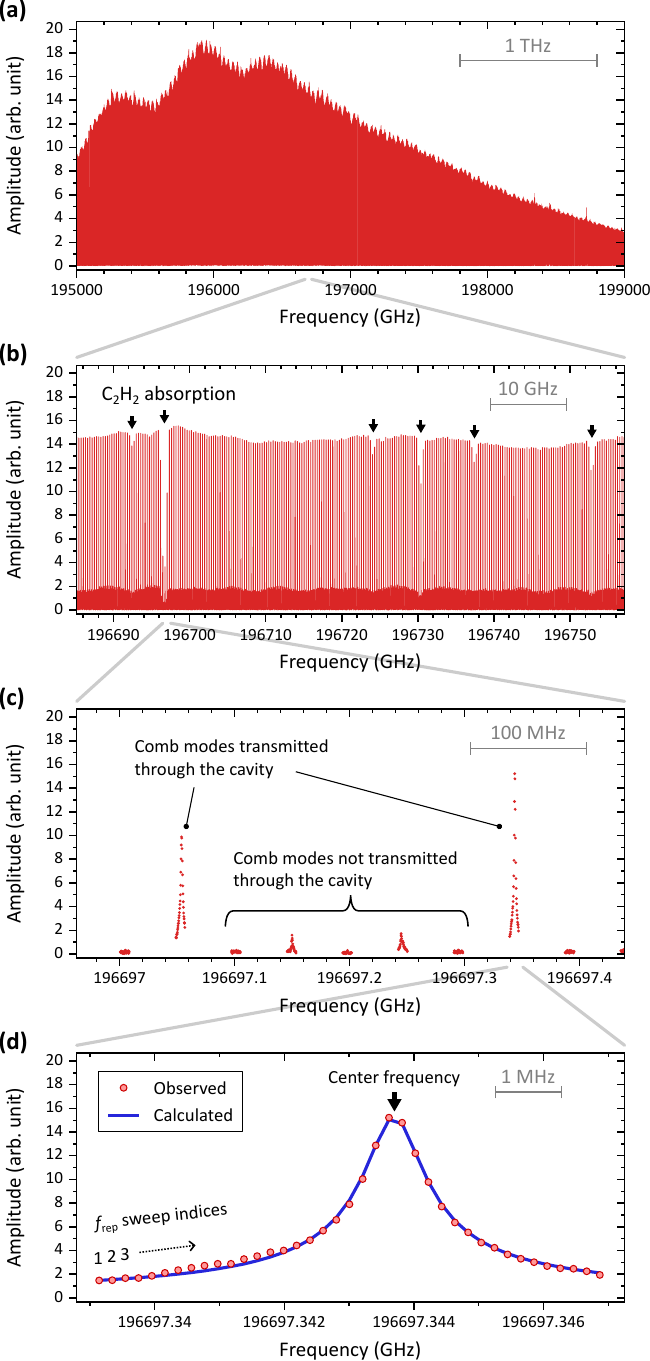}
\caption{
	Cavity transmitted spectrum observed by DCS. The frequency range is expanded from (a) to (d). (c) Each block of plots corresponds to the transmitted amplitude of the comb mode with the same mode number measured by sweeping the comb repetition rate. The two blocks represent the comb modes transmitted through the cavity, while the five blocks between them represent those not transmitted. This is because the cavity FSR is six times the comb repetition rate. (d) The red circle and the blue line, respectively, show the observed spectrum and the spectrum calculated by fitting the cavity transmittance function to the observed data.
}
\label{fig-3}
\end{figure}

Figure \ref{fig-3} shows the cavity transmission amplitude spectra measured by DCS. The frequency scale is progressively expanded from Fig. \ref{fig-3}(a) to \ref{fig-3}(d). Figure \ref{fig-3}(a) shows a broad spectrum covering almost the entire $\nu_1 + \nu_3$ band of \ce{^{12}C2H2}. In Fig. \ref{fig-3}(b), a large number of cavity transmission modes are observed, and absorption due to \ce{C2H2} can also be clearly identified. Figure \ref{fig-3}(c) shows the spectral shape of the cavity transmitted modes. Each plot corresponds to a comb mode with the same mode number, recorded by sweeping the repetition frequency and performing interleaved measurements. Of these plots, the two with large amplitudes correspond to comb modes that trace the transmission spectra of adjacent cavity modes. The five plots between them correspond to comb modes that do not resonate with the cavity. This is consistent with the fact that the cavity FSR is exactly six times the comb repetition frequency. The small transmission observed even for non-resonant comb modes (2nd and 4th plots) is attributed to the slight coupling of the comb beam to the transverse modes of the cavity. Figure \ref{fig-3}(d) shows an expanded spectrum of the right-side cavity transmitted mode of Fig. \ref{fig-3}(c). Each red circle represents the amplitude transmission spectrum measured by gradually changing the repetition frequency from the lower frequency side. The blue curve shows the spectrum calculated by a least-squares fit of the measured data (red circle) to the expression for the cavity amplitude transmission spectrum, given by the square root of a Lorentzian function,
\begin{equation}
f\left(\nu\right)=A\sqrt{\frac{\gamma^2}{\left(\nu-\nu_c\right)^2+\gamma^2}} \label{eq-5}
\end{equation}
where $\nu_\mathrm{c}$ is the center frequency, $\gamma$ is the half width at half maximum of a Lorentzian function of the power spectrum, and $A$ is the peak amplitude. These parameters are used in the fitting procedure to determine the center frequency $\nu_\mathrm{c}$.

\begin{figure}[htbp]
\centering\includegraphics[width=13cm]{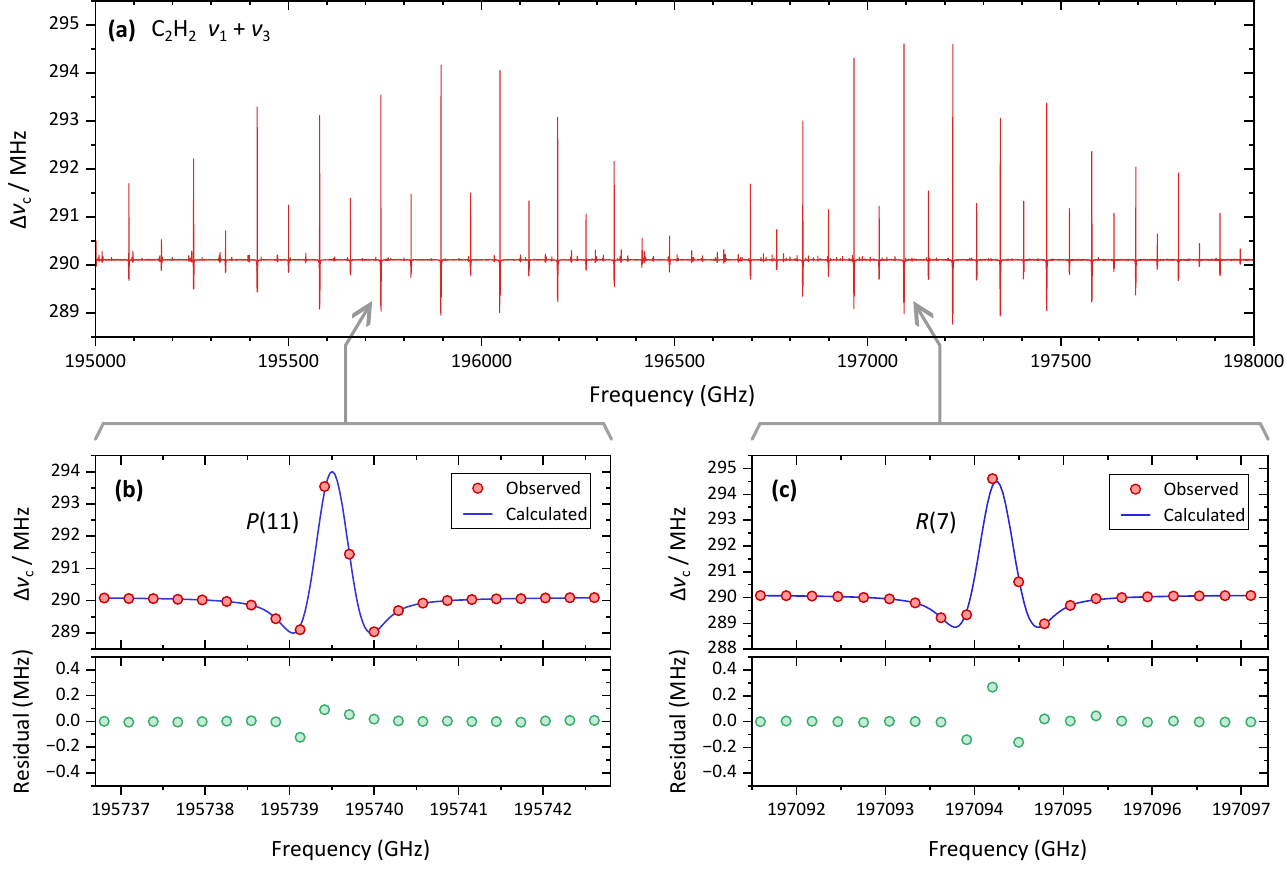}
\caption{
	CMD spectrum of \ce{C2H2}. (a) Broadband spectrum of \ce{^{12}C2H2} $\nu_1 + \nu_3$ band. $\Delta\nu_\mathrm{c}$: difference between the center frequencies of adjacent cavity modes. (b) and (c) show expanded spectra focused on the $P$(11) and $R$(7) transition. The red circle and the blue curve show the observed spectrum and the spectrum calculated by fitting the theoretical shape of the CMD spectrum to the observed data. The green circle shows the difference between the observed and calculated spectra.
}
\label{fig-4}
\end{figure}

We perform fitting in an identical way to that shown in Fig. \ref{fig-3}(d) for all observed cavity modes and determine the center frequency of each mode. Figure \ref{fig-4}(a) shows the CMD spectrum of \ce{^{12}C2H2} $\nu_1 + \nu_3$ band by calculating the differences between the center frequencies of adjacent cavity modes $\Delta\nu_\mathrm{c}$. The resulting spectrum is almost unaffected by the frequency dependence of the comb spectrum, as seen in Figs. \ref{fig-3}(a) and \ref{fig-3}(b), and has a noticeably flat background. A comparison with conventional DCS is discussed in the next section. The red circles in Figs. \ref{fig-4}(b) and \ref{fig-4}(c) show expanded spectra focusing on the $P$(11) and $R$(7) transitions, respectively. For the sample in the cavity cell in this study (\ce{C2H2} at a pressure of \qty{10}{Pa}), the Doppler-limited FWHM of the power absorption coefficient is calculated to be approximately \qty{470}{MHz} from the room temperature, molecular mass, and transition frequency. In contrast, the frequency spacing of the obtained CMD spectrum is approximately \qty{290}{MHz}, which corresponds to the FSR of the vacuum cavity. As a result, only a few data points contribute to a single absorption line, resulting in somewhat insufficient resolution for the sample spectrum. Consequently, in Fig. \ref{fig-4}(a), the intensity distribution of the transition band appears slightly different from the original shape of the $\nu_1 + \nu_3$ band. However, this will not be a problem in terms of the analysis of atmospheric gas components, since the transition linewidth of the gas will reach several GHz due to pressure broadening. In addition, it is possible to fill the gaps between data points and improve the resolution by changing the cavity length and performing interleaved measurements.

Finally, the pressure of \ce{C2H2} was determined by fitting the observed CMD spectrum to the model function. The model function $g(m)$ is based on Eq. (\ref{eq-4}) and can be expressed as follows:
\begin{align}
g\left(m\right)&=\nu\left(m+1\right)-\nu\left(m\right) \label{eq-6}\\
\nu\left(m\right)&=m\nu_\mathrm{FSR,vac}\left[1-\frac{c^2Sp}{4\pi\nu_0kT}V_\mathrm{disp}\left(\nu\left(m\right)-\nu_0\right)\right]^{-1} \label{eq-7}
\end{align}
Here, $m$ is the cavity mode number and is used as an independent variable for least-squares fitting, and $\nu(m)$ is the cavity mode frequency in the presence of absorption by the sample and is expressed in an implicit form as shown in Eq. (\ref{eq-7}). The blue curve in Fig. \ref{fig-4}(b) shows the calculated spectrum obtained by fitting the observed CMD spectrum to the model function given by Eqs. (\ref{eq-6}) and (\ref{eq-7}), using only p as a fitting parameter. The green circles indicate the difference between the calculated and observed spectra. In this analysis, the line intensity $S$ was taken from the HITRAN database \cite{GordonI2022}, $T$ was the room temperature of \qty{296}{K}, and $\nu_\mathrm{FSR,vac}$ was fixed at the average FSR determined in the frequency region without molecular transitions in Fig. \ref{fig-4}(a). The calculated spectra agree well with the observed spectra, and the difference between them is at most a few tens of kilohertz. In the \qty{195}{THz} to \qty{199}{THz} frequency range, partial spectra were extracted for each \ce{C2H2} transition with the line strength $S > \qty{e-21}{cm/molecule}$ within $\pm\qty{3}{GHz}$ of the transition frequency. Each partial spectrum was fitted to the model function to determine the pressure $p$ and its fitting error. The average pressure value, calculated with weights given by the square inverse of the fitting errors, was \qty{10.083(27)}{Pa}, where the value in parentheses represents the standard deviation. This result is consistent with the \ce{C2H2} pressure measured with a vacuum gauge (\qty{10}{Pa} with a nominal uncertainty of \qty{1}{Pa}). In situations such as measurements at atmospheric pressure, where more data points contribute to each transition line, the standard deviation is expected to decrease further.

\section{Discussions regarding measurement sensitivity}

As a metric of measurement sensitivity, we adopted the minimum detectable absorption coefficient (MDAC) \cite{OKeefeA1988}, defined by the fluctuation $\sigma_\alpha$ of the absorption coefficient $\alpha$. The MDAC was estimated by converting the fluctuation $\sigma_g$ of the CMD spectrum $g(m)$ into $\sigma_\alpha$. The standard deviation of the measured CMD spectrum in a frequency region without transition lines was $\sigma_g \simeq \qty{4}{kHz}$. The absorption coefficient $\alpha$ is given by Eq. (\ref{eq-3}), and the relationship between $\sigma_\alpha$ and $\sigma_g$ can be expressed, using pressure $p$ as an intermediate variable, as
\begin{equation}
\sigma_\alpha=\left|\frac{\partial\alpha}{\partial g}\right|\sigma_g=\left|\frac{\partial\alpha/\partial p}{\partial g/\partial p}\right|\sigma_g \label{eq-8}
\end{equation}
At the transition frequency (line center), where both $g(m)$ and $\alpha(\nu)$ reach their maxima, we numerically evaluated $\partial g/\partial p$ near $p = \qty{0}{Pa}$ from Eqs. (\ref{eq-6}) and (\ref{eq-7}) and $\partial\alpha/\partial p$ near $p = \qty{0}{Pa}$ from Eq. (\ref{eq-3}). This yielded $\sigma_\alpha \simeq \qty{1.4e-6}{cm^{-1}}$. In CMDS, the measurement sensitivity is, in principle, inversely proportional to the cavity finesse. If the finesse were increased from 300 (used in this study) to \num{20000}, which is comparable to that used in previous DC-CMDS and dual-comb CRDS studies \cite{CharczunD2022, LisakD2022}, the MDAC would be expected to improve to \qty{2.1e-8}{cm^{-1}} under a simple scaling assumption. This value is comparable to the MDACs inferred from previous studies of DC-CMDS and DC-CRDS, approximately \qty{2e-8}{cm^{-1}} and \qty{6e-9}{cm^{-1}}, respectively. Considering that those studies employed modulator-based frequency combs, which offer a higher comb mode power and thus a favorable SNR, it is noteworthy that high sensitivity is still expected for broadband spectroscopy using mode-locked fiber lasers, as in the present work. Sensitivity can also be improved by moving to the mid-infrared region, where molecular fundamental vibrational bands exhibit a stronger transition. For example, for \ce{^{12}C2H2}, the $\nu_3$ band (center wavelength: $\sim \qty{3.04}{\micro m}$) has a transition strength approximately 20 times larger than the $\nu_1 + \nu_3$ band (center wavelength: $\sim \qty{1.53}{\micro m}$) used in this study \cite{GordonI2022}, suggesting a corresponding improvement in sensitivity. In this respect as well, optical frequency combs based on mode-locked fiber lasers, which can be readily broadened or wavelength converted, are advantageous.

\begin{figure}[htbp]
\centering\includegraphics[width=13cm]{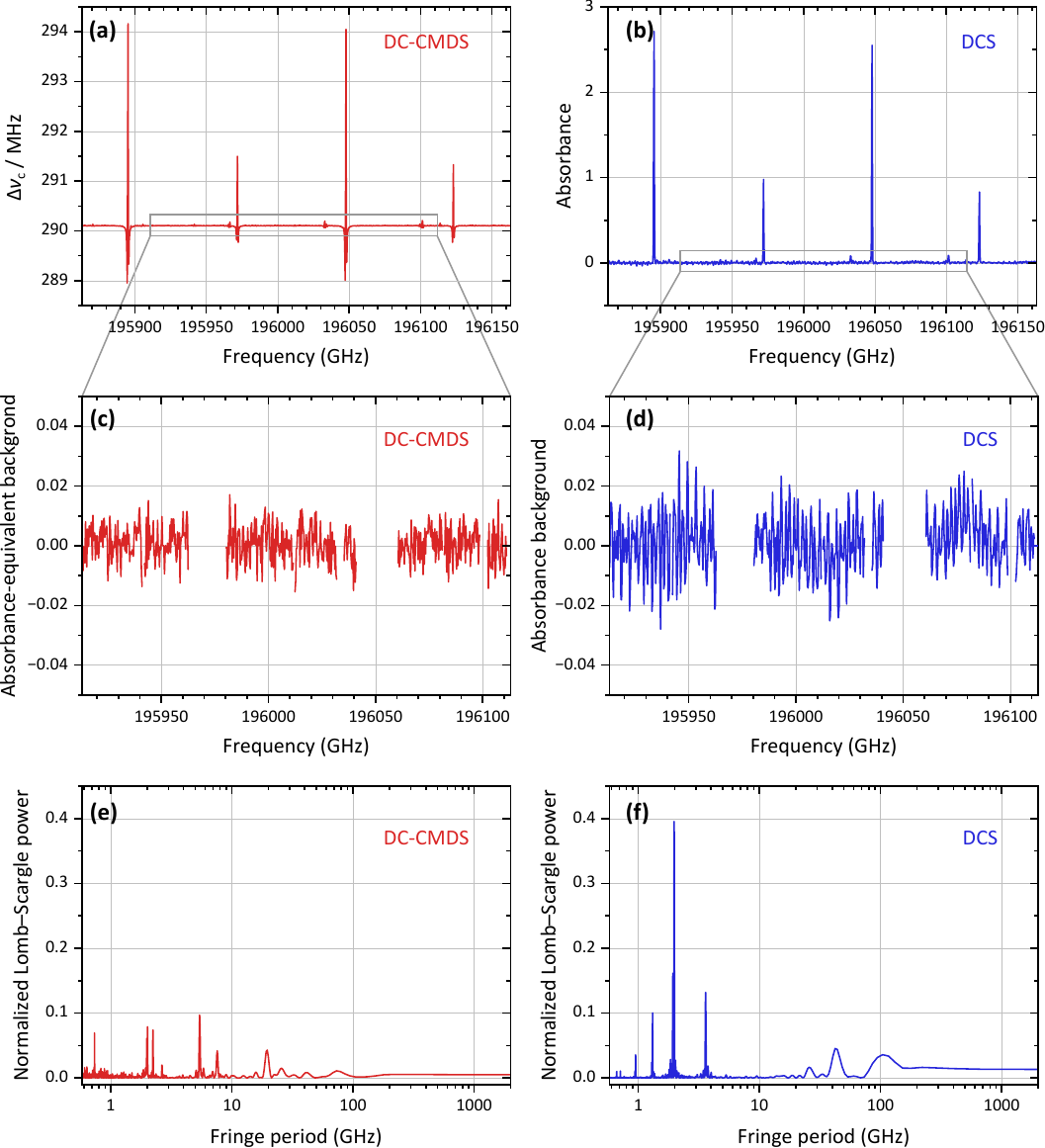}
\caption{
	Comparison of the backgrounds of DC-CMDS in this study and conventional DCS in Refs. \cite{IwakuniK2016b, OkuboS2017}. (a) and (b) show the spectra measured with DC-CMDS and DCS in the same frequency range. (c) and (d) show the background signals of (a) and (b), where the molecular transition is masked. The vertical axis of (c) is converted to the equivalent value of the absorbance. (e) and (f) show the normalized Lomb-Scargle powers of (c) and (d) to identify the spectral fringe components.
}
\label{fig-5}
\end{figure}

In addition, fringes in the background signal often limit the measurement sensitivity. Figure 5 compares the background in the spectrum of DC-CMDS obtained in this study with that in the spectrum of conventional DCS. Figures \ref{fig-5}(a) and \ref{fig-5}(b), respectively, show the \ce{C2H2} spectra of DC-CMDS and conventional DCS measured over the same frequency range. The spectrum of conventional DCS was taken from Refs. \cite{IwakuniK2016b, OkuboS2017}, where \ce{C2H2} at \qty{60}{Pa} was measured in a \qty{3}{m} White cell using a pair of frequency combs identical to those used in this study. The spectral frequency step in the conventional DCS was $f_\mathrm{rep,S} \simeq \qty{48.35}{MHz}$; to match the frequency step of the present DC-CMDS ($6f_\mathrm{rep} \simeq \qty{290}{MHz}$), the DCS data was averaged over every six consecutive points. Spectra with and without \ce{C2H2} in the cell, $I(\nu)$ and $I_0(\nu)$, respectively, were measured, and the absorbance spectrum was calculated as $\alpha(\nu) = -\ln \left[(I(\nu)/I_0(\nu)\right]$. Figures \ref{fig-5}(c) and \ref{fig-5}(d) compare the background signals after masking out the spectral regions containing molecular transitions. The boxed regions in Figs. \ref{fig-5}(a) and \ref{fig-5}(b) are enlarged to facilitate comparison. In Fig. \ref{fig-5}(c), the background in the CMD spectrum was converted to an equivalent absorbance using $\left|\partial\alpha/\partial g\right|$ in Eq. (\ref{eq-8}) and the effective optical path length of the cavity, $L_\mathrm{eff} = L/(1 - R)$ \cite{BaerD2002}. Because the measurement conditions (e.g., the average number of interferograms and the number of signal comb modes contributing to the measurement) differ for DC-CMDS and DCS, a direct comparison of the random-noise levels is not straightforward. However, in the spectrum obtained by DCS, residual fringes attributed to an etalon fringe superimposed on the comb spectral envelope remain with a peak-to-peak amplitude of up to $\sim 0.04$. To identify the fringe components more quantitatively, we performed a periodicity analysis of the background using the Lomb-Scargle method \cite{LombN1976, ScargleJ1982}. Figures \ref{fig-5}(e) and \ref{fig-5}(f) show periodograms of the normalized Lomb-Scargle power $P_\mathrm{LS}(\Lambda)$, calculated from the background signals in Figs. \ref{fig-5}(c) and \ref{fig-5}(d), respectively. Here, $P_\mathrm{LS}(\Lambda)$ is defined as the relative reduction in the residual sum of squares obtained by fitting a sinusoidal model (denoted below) to the background data (yielding $\chi^2(\Lambda)$), relative to that obtained by fitting a constant model to the same data (yielding $\chi_\mathrm{ref}^2$), and expressed as
\begin{equation}
P_\mathrm{LS}\left(\Lambda\right)=\frac{\chi_\mathrm{ref}^2-\chi^2\left(\Lambda\right)}{\chi_\mathrm{ref}^2} \label{eq-9}
\end{equation}
The sinusoidal model is expressed as
\begin{equation}
M\left(x;\Lambda\right)=A\cos\left(\frac{2\pi x}{\Lambda}\right)+B\sin\left(\frac{2\pi x}{\Lambda}\right)+C \label{eq-10}
\end{equation}
where $A$, $B$, and $C$ are fitting parameters, $x$ is an independent variable (optical frequency in this case), and $\Lambda$ is the Fourier period (i.e., the inverse of the Fourier frequency). A larger $P_\mathrm{LS}(\Lambda)$ indicates that the sinusoidal model is more appropriate and thus suggests the presence of fringe components. Whereas a fringe component with a period of approximately \qty{2}{GHz} was detected in the DCS background, the DC-CMDS background exhibits only a few weak peaks, indicating that fringe components are largely suppressed. Note that the normalization method used in DCS, namely dividing the spectrum measured with the sample by that measured without the sample, generally yields smaller residual fringes than approaches based on estimating the background spectrum. Nevertheless, the fringes cannot be completely removed, for example, due to temporal variations between the two measurements (with and without the sample). In summary, DC-CMDS enables the acquisition of spectra with negligible fringe components without special normalization and is therefore also superior to conventional DCS in terms of measurement sensitivity.

\section{Conclusion}

We developed a coherent dual-comb system that combines a mode-locked fiber laser with an optical cavity whose length can be tuned, enabling broadband CMDS with a spectral coverage expanded by a factor of about hundred beyond what has been reported to date. As a result, we obtained the entire transition spectrum of the \ce{^{12}C2H2} $\nu_1 + \nu_3$ band and confirmed that the background contribution was substantially reduced compared with conventional DCS without spectral normalization. By analyzing the obtained spectrum using a model function for the CMD spectrum, we confirmed that the model accurately reproduced the experimental data and determined the sample gas pressure with a relative standard deviation of \qty{0.27}{\%}. Furthermore, we evaluated the minimum detectable absorption coefficient (MDAC) to be \qty{1.4e-6}{cm^{-1}} from the fluctuation of the measured spectrum. Because the MDAC depends on the cavity finesse, further improvement is expected. In this study, we used a cavity with a finesse of 300; however, higher-finesse cavities of the order of \num{100000} are also feasible, indicating significant potential for sensitivity enhancement. The spectral range can be further extended by broadening the comb spectrum and by using cavity mirrors with a broad high-reflectance bandwidth.

The suppression of spectral background components demonstrated with our method is particularly effective for applications requiring a precise spectral analysis. In addition to the intrinsic advantages of DCS on the frequency (horizontal) axis namely a high resolution, a wide observable wavelength range, and a high frequency accuracy traceable to frequency standards, our method also improves performance on the amplitude (vertical) axis. By reducing spectral distortion, it enables, for example, a more accurate determination of molecular parameters, such as molecular constants, transition dipole moments, and pressure-shift and pressure-broadening coefficients, as well as more sensitive gas detection. Together with the high-speed measurement capability of DCS, this method is expected to contribute to expanding the range of DCS applications.

\begin{backmatter}

\bmsection{Funding}
Japan Society for the Promotion of Science (JSPS) KAKENHI Grant Number JP21H01852.

\bmsection{Disclosures}
The authors declare no conflicts of interest.

\bmsection{Data availability}
Data underlying the results presented in this paper are not publicly available at this time but may be obtained from the authors upon reasonable request.

\end{backmatter}



\bibliography{references}

@Article{Coddington2008,
  author           = {I. Coddington and W. C. Swann and N. R. Newbury},
  journal          = {Phys. Rev. Lett.},
  title            = {Coherent Multiheterodyne Spectroscopy Using Stabilized Optical Frequency Combs},
  year             = {2008},
  number           = {1},
  pages            = {013902},
  volume           = {100},
  doi              = {10.1103/PhysRevLett.100.013902},
  modificationdate = {2021-10-07T18:10:44},
}

@Article{Coddington2009b,
  author    = {I. Coddington and W. C. Swann and N. R. Newbury},
  journal   = {Opt. Lett.},
  title     = {Coherent linear optical sampling at 15 bits of resolution},
  year      = {2009},
  month     = {Jul},
  number    = {14},
  pages     = {2153--2155},
  volume    = {34},
  doi       = {10.1364/OL.34.002153},
  publisher = {OSA},
  url       = {http://ol.osa.org/abstract.cfm?URI=ol-34-14-2153},
}

@Article{Diddams2007,
  author  = {S. A. Diddams and L. Hollberg and V. Mbele},
  journal = {Nature},
  title   = {Molecular fingerprinting with the resolved modes of a femtosecond laser frequency comb},
  year    = {2007},
  pages   = {627--630},
  volume  = {445},
}

@Article{Fleisher2014,
  author  = {Fleisher, Adam J. and Bjork, Bryce J. and Bui, Thinh Q. and Cossel, Kevin C. and Okumura, Mitchio and Ye, Jun},
  journal = {J. Phys. Chem. Lett.},
  title   = {Mid-Infrared Time-Resolved Frequency Comb Spectroscopy of Transient Free Radicals},
  year    = {2014},
  number  = {13},
  pages   = {2241--2246},
  volume  = {5},
  doi     = {10.1021/jz5008559},
  eprint  = {http://dx.doi.org/10.1021/jz5008559},
  url     = {http://dx.doi.org/10.1021/jz5008559},
}

@Article{Hall2006,
  author    = {Hall, John L.},
  journal   = {Rev. Mod. Phys.},
  title     = {Nobel Lecture: Defining and measuring optical frequencies},
  year      = {2006},
  month     = {Nov},
  pages     = {1279--1295},
  volume    = {78},
  doi       = {10.1103/RevModPhys.78.1279},
  issue     = {4},
  publisher = {American Physical Society},
  url       = {http://link.aps.org/doi/10.1103/RevModPhys.78.1279},
}

@Article{Hansch2006,
  author    = {H\"{a}nsch, Theodor W.},
  journal   = {Rev. Mod. Phys.},
  title     = {Nobel Lecture: Passion for precision},
  year      = {2006},
  month     = {Nov},
  pages     = {1297--1309},
  volume    = {78},
  doi       = {10.1103/RevModPhys.78.1297},
  issue     = {4},
  publisher = {American Physical Society},
  url       = {http://link.aps.org/doi/10.1103/RevModPhys.78.1297},
}

@Article{KeilmannF2004,
  author           = {Fritz Keilmann and Christoph Gohle and Ronald Holzwarth},
  journal          = {Opt. Lett.},
  title            = {Time-domain mid-infrared frequency-comb spectrometer},
  year             = {2004},
  month            = {jul},
  number           = {13},
  pages            = {1542--1544},
  volume           = {29},
  doi              = {10.1364/ol.29.001542},
  modificationdate = {2021-10-07T17:47:17},
  publisher        = {The Optical Society},
  url              = {http://dx.doi.org/10.1364/OL.29.001542},
}

@Article{KloseA2016,
  author       = {Klose, Andrew and Ycas, Gabriel and Cruz, Flavio C. and Maser, Daniel L. and Diddams, Scott A.},
  journal      = {Appl. Phys. B},
  title        = {Rapid, broadband spectroscopic temperature measurement of {CO}$_2$ using {VIPA} spectroscopy},
  year         = {2016},
  issn         = {1432-0649},
  month        = Mar,
  number       = {4},
  pages        = {78},
  volume       = {122},
  doi          = {10.1007/s00340-016-6349-4},
  journaltitle = {Appl. Phys B},
  publisher    = {Springer Science and Business Media LLC},
  url          = {http://dx.doi.org/10.1007/s00340-016-6349-4},
}

@Article{Mandon2009,
  author  = {J. Mandon and G. Guelachvili and N. Picqu\'{e}},
  journal = {Nature Photon.},
  title   = {Fourier transform spectroscopy with a laser frequency comb},
  year    = {2009},
  pages   = {99--102},
  volume  = {3},
}

@Article{Minoshima2000,
  author    = {Kaoru Minoshima and Hirokazu Matsumoto},
  journal   = {Appl. Opt.},
  title     = {High-Accuracy Measurement of 240-m Distance in an Optical Tunnel by use of a Compact Femtosecond Laser},
  year      = {2000},
  month     = {Oct},
  number    = {30},
  pages     = {5512--5517},
  volume    = {39},
  doi       = {10.1364/AO.39.005512},
  publisher = {OSA},
  url       = {http://ao.osa.org/abstract.cfm?URI=ao-39-30-5512},
}

@Article{Okubo2015a,
  author    = {Sho Okubo and Kana Iwakuni and Hajime Inaba and Kazumoto Hosaka and Atsushi Onae and Hiroyuki Sasada and Feng-Lei Hong},
  journal   = {Appl. Phys. Express},
  title     = {Ultra-broadband dual-comb spectroscopy across 1.0--1.9 \textmu m},
  year      = {2015},
  issn      = {1882-0786},
  month     = {July},
  number    = {8},
  pages     = {082402},
  volume    = {8},
  doi       = {10.7567/APEX.8.082402},
  publisher = {IOP Publishing},
  url       = {http://stacks.iop.org/1882-0786/8/i=8/a=082402},
}

@Article{Okubo2015d,
  author    = {Sho Okubo and Yi-Da Hsieh and Hajime Inaba and Atsushi Onae and Mamoru Hashimoto and Takeshi Yasui},
  journal   = {Opt. Express},
  title     = {Near-infrared broadband dual-frequency-comb spectroscopy with a resolution beyond the Fourier limit determined by the observation time window},
  year      = {2015},
  month     = {Dec},
  number    = {26},
  pages     = {33184--33193},
  volume    = {23},
  doi       = {10.1364/OE.23.033184},
  publisher = {OSA},
  url       = {http://www.opticsexpress.org/abstract.cfm?URI=oe-23-26-33184},
}

@Article{Rieker2014,
  author    = {G. B. Rieker and F. R. Giorgetta and W. C. Swann and J. Kofler and A. M. Zolot and L. C. Sinclair and E. Baumann and C. Cromer and G. Petron and C. Sweeney and P. P. Tans and I. Coddington and N. R. Newbury},
  journal   = {Optica},
  title     = {Frequency-comb-based remote sensing of greenhouse gases over kilometer air paths},
  year      = {2014},
  month     = {Nov},
  number    = {5},
  pages     = {290--298},
  volume    = {1},
  doi       = {10.1364/OPTICA.1.000290},
  publisher = {OSA},
  url       = {http://www.osapublishing.org/optica/abstract.cfm?URI=optica-1-5-290},
}

@Article{Schiller2002,
  author    = {S. Schiller},
  journal   = {Opt. Lett.},
  title     = {Spectrometry with frequency combs},
  year      = {2002},
  month     = {May},
  number    = {9},
  pages     = {766--768},
  volume    = {27},
  doi       = {10.1364/OL.27.000766},
  publisher = {OSA},
  url       = {http://ol.osa.org/abstract.cfm?URI=ol-27-9-766},
}

@Article{Yamada2009,
  author    = {Koichi M.T. Yamada and Atsushi Onae and Feng-Lei Hong and Hajime Inaba and Tadao Shimizu},
  journal   = {C. R. Phys.},
  title     = {Precise determination of the Doppler width of a rovibrational absorption line using a comb-locked diode laser},
  year      = {2009},
  issn      = {1631-0705},
  month     = Nov,
  number    = {9},
  pages     = {907--915},
  volume    = {10},
  doi       = {10.1016/j.crhy.2009.10.011},
  publisher = {MathDoc/Centre Mersenne},
  url       = {http://www.sciencedirect.com/science/article/pii/S1631070509001571},
}

@Article{McCrackenR2017,
  author    = {Richard A. McCracken and Jake M. Charsley and Derryck T. Reid},
  journal   = {Opt. Express},
  title     = {A decade of astrocombs: recent advances in frequency combs for astronomy},
  year      = {2017},
  month     = {jun},
  number    = {13},
  pages     = {15058--15078},
  volume    = {25},
  doi       = {10.1364/oe.25.015058},
  publisher = {The Optical Society},
}

@Article{OkuboS2017,
  author           = {Sho Okubo and Kana Iwakuni and Koichi M.T. Yamada and Hajime Inaba and Atsushi Onae and Feng-Lei Hong and Hiroyuki Sasada},
  journal          = {J. Mol. Spectrosc.},
  title            = {Transition dipole-moment of the $\nu_{1} + \nu_{3}$ band of acetylene measured with dual-comb Fourier-transform spectroscopy},
  year             = {2017},
  month            = {nov},
  pages            = {10--16},
  volume           = {341},
  doi              = {10.1016/j.jms.2017.09.001},
  modificationdate = {2021-10-07T17:32:53},
  publisher        = {Elsevier {BV}},
}

@Article{CichM2013,
  author    = {Matthew J. Cich and Damien Forthomme and Christopher P. McRaven and Gary V. Lopez and Gregory E. Hall and Trevor J. Sears and Arlan. W. Mantz},
  journal   = {J. Phys. Chem. A},
  title     = {Temperature-Dependent, Nitrogen-Perturbed Line Shape Measurements in the $\nu_{1} + \nu_{3}$ Band of Acetylene Using a Diode Laser Referenced to a Frequency Comb},
  year      = {2013},
  month     = {dec},
  number    = {50},
  pages     = {13908--13918},
  volume    = {117},
  doi       = {10.1021/jp408960e},
  publisher = {American Chemical Society ({ACS})},
}

@Article{WaxmanE2017,
  author    = {Eleanor M. Waxman and Kevin C. Cossel and Gar-Wing Truong and Fabrizio R. Giorgetta and William C. Swann and Sean Coburn and Robert J. Wright and Gregory B. Rieker and Ian Coddington and Nathan R. Newbury},
  journal   = {Atmos. Meas. Tech.},
  title     = {Intercomparison of open-path trace gas measurements with two dual-frequency-comb spectrometers},
  year      = {2017},
  month     = {sep},
  number    = {9},
  pages     = {3295--3311},
  volume    = {10},
  doi       = {10.5194/amt-10-3295-2017},
  publisher = {Copernicus {GmbH}},
}

@Article{EndoM2018,
  author    = {Mamoru Endo and Tyko D. Shoji and Thomas R. Schibli},
  journal   = {IEEE J. Sel. Topics Quantum Electron.},
  title     = {Ultralow Noise Optical Frequency Combs},
  year      = {2018},
  month     = {sep},
  number    = {5},
  pages     = {1--13},
  volume    = {24},
  doi       = {10.1109/jstqe.2018.2818461},
  publisher = {Institute of Electrical and Electronics Engineers ({IEEE})},
}

@Article{CyganA2016,
  author    = {A. Cygan and S. W{\'{o}}jtewicz and G. Kowzan and M. Zaborowski and P. Wcis{\l}o and J. Nawrocki and P. Krehlik and {\L}. {\'{S}}liwczy{\'{n}}ski and M. Lipi{\'{n}}ski and P. Mas{\l}owski and R. Ciury{\l}o and D. Lisak},
  journal   = {J. Chem. Phys.},
  title     = {Absolute molecular transition frequencies measured by three cavity-enhanced spectroscopy techniques},
  year      = {2016},
  month     = {jun},
  number    = {21},
  pages     = {214202},
  volume    = {144},
  doi       = {10.1063/1.4952651},
  publisher = {{AIP} Publishing},
}

@Article{MurphyM2007,
  author    = {M. T. Murphy and Th. Udem and R. Holzwarth and A. Sizmann and L. Pasquini and C. Araujo-Hauck and H. Dekker and S. D'Odorico and M. Fischer and T. W. H\"{a}nsch and A. Manescau},
  journal   = {Mon. Not. R. Astron. Soc.},
  title     = {High-precision wavelength calibration of astronomical spectrographs with laser frequency combs},
  year      = {2007},
  month     = {aug},
  number    = {2},
  pages     = {839--847},
  volume    = {380},
  doi       = {10.1111/j.1365-2966.2007.12147.x},
  publisher = {Oxford University Press ({OUP})},
}

@Article{IwakuniK2019,
  author    = {K. Iwakuni and T. Q. Bui and J. F. Niedermeyer and T. Sukegawa and J. Ye},
  journal   = {Opt. Express},
  title     = {Comb-resolved spectroscopy with immersion grating in long-wave infrared},
  year      = {2019},
  month     = {jan},
  number    = {3},
  pages     = {1911--1921},
  volume    = {27},
  doi       = {10.1364/oe.27.001911},
  publisher = {The Optical Society},
}

@Article{PicqueN2019,
  author    = {Nathalie Picqu{\'{e}} and Theodor W. Hänsch},
  journal   = {Nat. Photonics},
  title     = {Frequency comb spectroscopy},
  year      = {2019},
  month     = {feb},
  number    = {3},
  pages     = {146--157},
  volume    = {13},
  doi       = {10.1038/s41566-018-0347-5},
  publisher = {Springer Science and Business Media {LLC}},
}

@Article{CyganA2015,
  author    = {Agata Cygan and Piotr Wcis{\l}o and Szymon W{\'{o}}jtewicz and Piotr Mas{\l}owski and Joseph T. Hodges and Roman Ciury{\l}o and Daniel Lisak},
  journal   = {Opt. Express},
  title     = {One-dimensional frequency-based spectroscopy},
  year      = {2015},
  month     = {may},
  number    = {11},
  pages     = {14472--14486},
  volume    = {23},
  doi       = {10.1364/oe.23.014472},
  publisher = {The Optical Society},
}

@Article{OKeefeA1988,
  author    = {Anthony O'Keefe and David A. G. Deacon},
  journal   = {Rev. Sci. Instrum.},
  title     = {Cavity ring-down optical spectrometer for absorption measurements using pulsed laser sources},
  year      = {1988},
  month     = {dec},
  number    = {12},
  pages     = {2544--2551},
  volume    = {59},
  doi       = {10.1063/1.1139895},
  publisher = {{AIP} Publishing},
}

@Article{ColeR2019,
  author    = {Ryan K. Cole and Amanda S. Makowiecki and Nazanin Hoghooghi and Gregory B. Rieker},
  journal   = {Opt. Express},
  title     = {Baseline-free quantitative absorption spectroscopy based on cepstral analysis},
  year      = {2019},
  month     = {dec},
  number    = {26},
  pages     = {37920--37939},
  volume    = {27},
  doi       = {10.1364/oe.27.037920},
  publisher = {The Optical Society},
}

@Article{BaerD2002,
  author    = {D.S. Baer and J.B. Paul and M. Gupta and A. O{\textquotesingle}Keefe},
  journal   = {Appl. Phys. B: Lasers Opt.},
  title     = {Sensitive absorption measurements in the near-infrared region using off-axis integrated-cavity-output spectroscopy},
  year      = {2002},
  month     = {sep},
  number    = {2-3},
  pages     = {261--265},
  volume    = {75},
  doi       = {10.1007/s00340-002-0971-z},
  publisher = {Springer Science and Business Media {LLC}},
}

@Article{CharczunD2022,
  author    = {D. Charczun and A. Nishiyama and G. Kowzan and A. Cygan and T. Voumard and T. Wildi and T. Herr and V. Brasch and D. Lisak and P. Mas{\l}owski},
  journal   = {Measurement},
  title     = {Dual-comb cavity-mode width and shift spectroscopy},
  year      = {2022},
  month     = {jan},
  pages     = {110519},
  volume    = {188},
  doi       = {10.1016/j.measurement.2021.110519},
  publisher = {Elsevier {BV}},
}

@Article{GordonI2022,
  author    = {I.E. Gordon and L.S. Rothman and R.J. Hargreaves and R. Hashemi and E.V. Karlovets and F.M. Skinner and E.K. Conway and C. Hill and R.V. Kochanov and Y. Tan and P. Wcis{\l}o and A.A. Finenko and K. Nelson and P.F. Bernath and M. Birk and V. Boudon and A. Campargue and K.V. Chance and A. Coustenis and B.J. Drouin and J.{\textendash}M. Flaud and R.R. Gamache and J.T. Hodges and D. Jacquemart and E.J. Mlawer and A.V. Nikitin and V.I. Perevalov and M. Rotger and J. Tennyson and G.C. Toon and H. Tran and V.G. Tyuterev and E.M. Adkins and A. Baker and A. Barbe and E. Can{\`{e}} and A.G. Cs{\'{a}}sz{\'{a}}r and A. Dudaryonok and O. Egorov and A.J. Fleisher and H. Fleurbaey and A. Foltynowicz and T. Furtenbacher and J.J. Harrison and J.{\textendash}M. Hartmann and V.{\textendash}M. Horneman and X. Huang and T. Karman and J. Karns and S. Kassi and I. Kleiner and V. Kofman and F. Kwabia{\textendash}Tchana and N.N. Lavrentieva and T.J. Lee and D.A. Long and A.A. Lukashevskaya and O.M. Lyulin and V.Yu. Makhnev and W. Matt and S.T. Massie and M. Melosso and S.N. Mikhailenko and D. Mondelain and H.S.P. Müller and O.V. Naumenko and A. Perrin and O.L. Polyansky and E. Raddaoui and P.L. Raston and Z.D. Reed and M. Rey and C. Richard and R. T{\'{o}}bi{\'{a}}s and I. Sadiek and D.W. Schwenke and E. Starikova and K. Sung and F. Tamassia and S.A. Tashkun and J. Vander Auwera and I.A. Vasilenko and A.A. Vigasin and G.L. Villanueva and B. Vispoel and G. Wagner and A. Yachmenev and S.N. Yurchenko},
  journal   = {J. Quant. Spectrosc. Radiat. Transf.},
  title     = {The {HITRAN}2020 molecular spectroscopic database},
  year      = {2022},
  month     = {jan},
  pages     = {107949},
  volume    = {277},
  doi       = {10.1016/j.jqsrt.2021.107949},
  publisher = {Elsevier {BV}},
}

@Article{IwakuniK2016b,
  author           = {Iwakuni, Kana and Okubo, Sho and Yamada, Koichi M. T. and Inaba, Hajime and Onae, Atsushi and Hong, Feng-Lei and Sasada, Hiroyuki},
  journal          = {Phys. Rev. Lett.},
  title            = {Ortho-Para-Dependent Pressure Effects Observed in the Near Infrared Band of Acetylene by Dual-Comb Spectroscopy},
  year             = {2016},
  month            = {Sep},
  pages            = {143902},
  volume           = {117},
  doi              = {10.1103/PhysRevLett.117.143902},
  issue            = {14},
  modificationdate = {2021-10-07T17:25:21},
  numpages         = {5},
  publisher        = {American Physical Society},
  url              = {http://link.aps.org/doi/10.1103/PhysRevLett.117.143902},
}

@Article{ShimizuY2018,
  author    = {Yukiko Shimizu and Sho Okubo and Atsushi Onae and Koichi M. T. Yamada and Hajime Inaba},
  journal   = {Appl. Phys. B},
  title     = {Molecular gas thermometry on acetylene using dual-comb spectroscopy: analysis of rotational energy distribution},
  year      = {2018},
  month     = {apr},
  number    = {4},
  pages     = {71},
  volume    = {124},
  doi       = {10.1007/s00340-018-6933-x},
  publisher = {Springer Nature},
}

@Article{NakamuraK2023,
  author    = {Keisuke Nakamura and Ken Kashiwagi and Sho Okubo and Hajime Inaba},
  journal   = {Opt. Express},
  title     = {Erbium-doped-fiber-based broad visible range frequency comb with a 30 {GHz} mode spacing for astronomical applications},
  year      = {2023},
  month     = {may},
  number    = {12},
  pages     = {20274--20285},
  volume    = {31},
  doi       = {10.1364/oe.487279},
  publisher = {Optica Publishing Group},
}

@Article{OkuboS2023,
  author    = {Sho Okubo and Kana Iwakuni and Hideki Kato and Feng-Lei Hong and Hiroyuki Sasada and Hajime Inaba and Koichi M.T. Yamada},
  journal   = {J. Mol. Spectrosc.},
  title     = {The pressure effect on the line profiles observed in the $\nu_{1} + \nu_{3}$ band of acetylene: Revisited},
  year      = {2023},
  month     = {aug},
  pages     = {111823},
  volume    = {396},
  doi       = {10.1016/j.jms.2023.111823},
  publisher = {Elsevier {BV}},
}

@Article{HanJ2024,
  author    = {Han, Jin-Jian and Zhong, Wei and Zhao, Ruo-Can and Zeng, Ting and Li, Min and Lu, Jian and Peng, Xin-Xin and Shi, Xi-Ping and Yin, Qin and Wang, Yong and Esamdin, Ali and Shen, Qi and Guan, Jian-Yu and Hou, Lei and Ren, Ji-Gang and Jia, Jian-Jun and Wang, Yu and Jiang, Hai-Feng and Xue, Xiang-Hui and Zhang, Qiang and Dou, Xian-Kang and Pan, Jian-Wei},
  journal   = {Nat. Photonics},
  title     = {Dual-comb spectroscopy over a 100 km open-air path},
  year      = {2024},
  issn      = {1749-4893},
  month     = sep,
  number    = {11},
  pages     = {1195--1202},
  volume    = {18},
  doi       = {10.1038/s41566-024-01525-9},
  publisher = {Springer Science and Business Media LLC},
}

@Article{LisakD2022,
  author       = {Lisak, Daniel and Charczun, Dominik and Nishiyama, Akiko and Voumard, Thibault and Wildi, Thibault and Kowzan, Grzegorz and Brasch, Victor and Herr, Tobias and Fleisher, Adam J. and Hodges, Joseph T. and Ciuryło, Roman and Cygan, Agata and Masłowski, Piotr},
  journal      = {Sci. Rep.},
  title        = {Dual-comb cavity ring-down spectroscopy},
  volume       = {12},
  year         = {2022},
  issn         = {2045-2322},
  month        = feb,
  number       = {1},
  pages        = {2377},
  doi          = {10.1038/s41598-022-05926-0},
  publisher    = {Springer Science and Business Media LLC},
}

@Article{CyganA2013,
  author    = {Cygan, Agata and Lisak, Daniel and Morzyński, Piotr and Bober, Marcin and Zawada, Michał and Pazderski, Eugeniusz and Ciuryło, Roman},
  journal   = {Opt. Express},
  title     = {Cavity mode-width spectroscopy with widely tunable ultra narrow laser},
  year      = {2013},
  issn      = {1094-4087},
  month     = nov,
  number    = {24},
  pages     = {29744--29754},
  volume    = {21},
  doi       = {10.1364/oe.21.029744},
  publisher = {Optica Publishing Group},
}

@Article{ParriauxA2022,
  author    = {Parriaux, Alexandre and Hammani, Kamal and Thomazo, Christophe and Musset, Olivier and Millot, Guy},
  journal   = {Phys. Rev. Research},
  title     = {Isotope ratio dual-comb spectrometer},
  year      = {2022},
  issn      = {2643-1564},
  month     = may,
  number    = {2},
  pages     = {023098},
  volume    = {4},
  doi       = {10.1103/physrevresearch.4.023098},
  publisher = {American Physical Society (APS)},
}

@Article{GattiD2013,
  author    = {Gatti, Davide and Mills, Andrew A. and De Vizia, Maria Domenica and Mohr, Christian and Hartl, Ingmar and Marangoni, Marco and Fermann, Martin and Gianfrani, Livio},
  journal   = {Phys. Rev. A},
  title     = {Frequency-comb-calibrated Doppler broadening thermometry},
  year      = {2013},
  issn      = {1094-1622},
  month     = jul,
  number    = {1},
  pages     = {012514},
  volume    = {88},
  doi       = {10.1103/physreva.88.012514},
  publisher = {American Physical Society (APS)},
}

@Article{OhJ2005,
  author    = {Oh, Jeong Seok and Kim, Seung-Woo},
  journal   = {Opt. Lett.},
  title     = {Femtosecond laser pulses for surface-profile metrology},
  year      = {2005},
  issn      = {1539-4794},
  month     = oct,
  number    = {19},
  pages     = {2650--2652},
  volume    = {30},
  doi       = {10.1364/ol.30.002650},
  publisher = {Optica Publishing Group},
}

@Article{HartmannJ2013,
  author    = {Hartmann, J.-M. and Tran, H. and Ngo, N. H. and Landsheere, X. and Chelin, P. and Lu, Y. and Liu, A.-W. and Hu, S.-M. and Gianfrani, L. and Casa, G. and Castrillo, A. and Lepère, M. and Delière, Q. and Dhyne, M. and Fissiaux, L.},
  journal   = {Phys. Rev. A},
  title     = {Ab initio calculations of the spectral shapes of {CO}$_{2}$ isolated lines including non-Voigt effects and comparisons with experiments},
  year      = {2013},
  issn      = {1094-1622},
  month     = jan,
  number    = {1},
  pages     = {013403},
  volume    = {87},
  doi       = {10.1103/physreva.87.013403},
  publisher = {American Physical Society (APS)},
}

@Article{WeerasekaraC2024,
  author    = {Weerasekara, Chinthaka and Morris, Lindsay C. and Malarich, Nathan A. and Giorgetta, Fabrizio R. and Herman, Daniel I. and Cossel, Kevin C. and Newbury, Nathan R. and Owensby, Clenton E. and Welch, Stephen M. and Blaga, Cosmin and DePaola, Brett D. and Coddington, Ian and Washburn, Brian R. and Santos, Eduardo A.},
  journal   = {Atmos. Meas. Tech.},
  title     = {Using open-path dual-comb spectroscopy to monitor methane emissions from simulated grazing cattle},
  year      = {2024},
  issn      = {1867-8548},
  month     = oct,
  number    = {20},
  pages     = {6107--6117},
  volume    = {17},
  doi       = {10.5194/amt-17-6107-2024},
  publisher = {Copernicus GmbH},
}

@Article{AlrahmanC2014,
  author    = {Alrahman, Chadi Abd and Khodabakhsh, Amir and Schmidt, Florian M. and Qu, Zhechao and Foltynowicz, Aleksandra},
  journal   = {Opt. Express},
  title     = {Cavity-enhanced optical frequency comb spectroscopy of high-temperature {H}$_2${O} in a flame},
  year      = {2014},
  issn      = {1094-4087},
  month     = may,
  number    = {11},
  pages     = {13889--13895},
  volume    = {22},
  doi       = {10.1364/oe.22.013889},
  publisher = {Optica Publishing Group},
}

@Article{ScargleJ1982,
  author    = {Scargle, J. D.},
  journal   = {Astrophys. J.},
  title     = {Studies in astronomical time series analysis. {II} - Statistical aspects of spectral analysis of unevenly spaced data},
  year      = {1982},
  issn      = {1538-4357},
  month     = dec,
  pages     = {835--853},
  volume    = {263},
  doi       = {10.1086/160554},
  publisher = {American Astronomical Society},
}

@Article{LombN1976,
  author    = {Lomb, N. R.},
  journal   = {Astrophys. Space Sci.},
  title     = {Least-squares frequency analysis of unequally spaced data},
  year      = {1976},
  issn      = {1572-946X},
  month     = feb,
  number    = {2},
  pages     = {447--462},
  volume    = {39},
  doi       = {10.1007/bf00648343},
  publisher = {Springer Science and Business Media LLC},
}

@Article{HashiguchiK2024,
  author    = {Hashiguchi, Koji and Amano, Minami and Cygan, Agata and Lisak, Daniel and Ciury\l o, Roman and Abe, Hisashi},
  journal   = {AIP Conf. Proc.},
  title     = {Improvement of the cavity in {CRDS} for high-precision measurement of trace moisture},
  year      = {2024},
  issn      = {0094-243X},
  pages     = {140004},
  volume    = {3230},
  booktitle = {TEMPERATURE: ITS MEASUREMENT AND CONTROL IN SCIENCE AND INDUSTRY, VOLUME 9: Proceedings of the Tenth International Temperature Symposium},
  doi       = {10.1063/5.0234527},
  publisher = {AIP Publishing},
}

@Article{RamondT2002,
  author    = {Ramond, T. M. and Diddams, S. A. and Hollberg, L. and Bartels, A.},
  journal   = {Opt. Lett.},
  title     = {Phase-coherent link from optical to microwave frequencies by means of the broadband continuum from a 1-{GHz} {Ti}:sapphire femtosecond oscillator},
  year      = {2002},
  issn      = {1539-4794},
  month     = oct,
  number    = {20},
  pages     = {1842--1844},
  volume    = {27},
  doi       = {10.1364/ol.27.001842},
  publisher = {Optica Publishing Group},
}

@Article{IrimatsugawaT2021,
  author           = {Tomoya Irimatsugawa and Yukiko Shimizu and Sho Okubo and Hajime Inaba},
  journal          = {Opt. Express},
  title            = {Cosine similarity for quantitatively evaluating the degree of change in an optical frequency comb spectra},
  year             = {2021},
  month            = {oct},
  number           = {22},
  pages            = {35613--35622},
  volume           = {29},
  doi              = {10.1364/oe.435679},
  modificationdate = {2021-10-16T17:08:53},
  publisher        = {The Optical Society},
}

@Article{Drever1983,
  author    = {R. W. P. Drever and J. L. Hall and F. V. Kowalski and J. Hough and G. M. Ford and A. J. Munley and H. Ward},
  journal   = {Appl. Phys. B},
  title     = {Laser phase and frequency stabilization using an optical resonator},
  year      = {1983},
  month     = {jun},
  number    = {2},
  pages     = {97--105},
  volume    = {31},
  doi       = {10.1007/BF00702605},
  publisher = {Springer Science and Business Media {LLC}},
}

@Article{Yasui2015a,
  author    = {Takeshi Yasui and Yuki Iyonaga and Yi-Da Hsieh and Yoshiyuki Sakaguchi and Francis Hindle and Shuko Yokoyama and Tsutomu Araki and Mamoru Hashimoto},
  journal   = {Optica},
  title     = {Super-resolution discrete Fourier transform spectroscopy beyond time-window size limitation using precisely periodic pulsed radiation},
  year      = {2015},
  month     = {May},
  number    = {5},
  pages     = {460--467},
  volume    = {2},
  doi       = {10.1364/OPTICA.2.000460},
  publisher = {OSA},
  url       = {http://www.osapublishing.org/optica/abstract.cfm?URI=optica-2-5-460},
}

@Article{FelderR2005,
  author       = {Felder, R},
  journal      = {Metrologia},
  pages        = {323--325},
  title        = {Practical realization of the definition of the metre, including recommended radiations of other optical frequency standards (2003)},
  volume       = {42},
  year         = {2005},
  issn         = {1681-7575},
  month        = Jul,
  number       = {4},
  doi          = {10.1088/0026-1394/42/4/018},
  publisher    = {IOP Publishing},
}

@Article{WangZ2024b,
  author       = {Wang, Zhen and Nie, Qinxue and Sun, Haojia and Wang, Qiang and Borri, Simone and De Natale, Paolo and Ren, Wei},
  journal      = {Light Sci. Appl.},
  pages        = {11},
  title        = {Cavity-enhanced photoacoustic dual-comb spectroscopy},
  volume       = {13},
  year         = {2024},
  issn         = {2047-7538},
  month        = jan,
  number       = {1},
  creationdate = {2025-11-25T20:12:04},
  doi          = {10.1038/s41377-023-01353-6},
  publisher    = {Springer Science and Business Media LLC},
}






\end{document}